# Improving position resolution of neutron detectors with ultra-thin B$_4$C foils

**N. F. V. Duarte[a], J. S. Marcoş[a], A. Antognini[b,c], C. Klauser[c], S. A. Felix[a], C. M. B. Monteiro[a] and F. D. Amaro[a,1]**

[a] *LIBPhys – Physics Department, University of Coimbra, 3004-516 Coimbra, Portugal*

[b] *Institute for Particle Physics and Astrophysics, ETH Zurich, 8093 Zurich, Switzerland*

[c] *Paul Scherrer Institute, 5232 Villigen, Switzerland*

*E-mail*: famaro@uc.pt

ABSTRACT: A new technique for detection of slow neutrons with gaseous detectors using ultra-thin layers with $^{10}$B atoms is presented. The reaction between a thermal neutron and a $^{10}$B atom releases 2 secondary particles, namely a $^7$Li ion and an alpha particle, which due to momentum conservation are emitted in opposite directions, along the same line (back to back). Current boron coated neutron detectors are equipped with $^{10}$B films with thicknesses of several micrometers, deposited on very thick substrate plates. However, since the ranges of the $^7$Li ion and the alpha particle are of few micrometeres in most materials, one of these particles is always lost in the $^{10}$B layer or substrate. As such, these detectors lose the ability to reconstruct the reaction line of action and to precisely determine the neutron position, as only one of the two secondary particles track can be measured. With the technique now presented, the sum of the $^{10}$B layer and the substrate thicknesses is small enough to allow for both secondary particles to escape and ionize the gas in opposite sides of the $^{10}$B converter foil. Independent readout structures, one on each side of the $^{10}$B converter foil, detect each secondary particle and determine its track centroid and the deposited energy. Since the two secondary particles are emitted back to back, the neutron position can be obtained by combining the information recorded by the two readout structures. Through GEANT4 simulations, we verified that the spatial resolution can be significantly improved: our results show that, by using a B$_4$C layer with a thickness of 1 μm on a 0.9 μm Mylar substrate, the spatial resolution can by improved by a factor of 8, compared to conventional detectors with thick $^{10}$B detection layers.

# Contents

## 1. Introduction

Neutrons properties turn them into excellent probes for the investigation of matter, and these are used in different scientific fields such as nuclear and particle physics [1], materials science [2], molecular dynamic studies [3] and crystallography [4], as well as for medical therapy [5] and homeland security applications [6]. Modern large-scale neutron facilities such as the U.S. Spallation Neutron Source (SNS), the China Spallation Neutron Source (CSNS) and the currently in construction European Spallation Source (ESS), depend on the deployment of position sensitive neutron detectors (PSND) with high performance in aspects such as counting rate capability and spatial resolution.

    Neutron detection is usually achieved via nuclear capture reactions, in which the neutron is absorbed by the nucleus of an atom, which then becomes unstable and decays into two highly ionizing charged particles. The cross-section to thermal neutrons of these reactions is particularly high for few isotopes: $^3$He (5333 barn), $^{10}$B (3837 barn) and $^6$Li (940 barn). Because the neutron capture cross-section scales with $1/\sqrt{E}$, where E is the neutron energy, neutrons used in irradiation experiments typically undergo a moderation process in which their energy is dissipated through successive elastic scattering interactions within a moderator medium, until thermal energy (0.025 eV) is reached.

## 1.1 $^3$He proportional counters

The neutron capture reaction by $^3$He atoms results in the emission of one triton and one proton with energies of 191 keV and 573 keV, respectively, accompanied by a 0.764 MeV gamma-ray. The fact that $^3$He is naturally a stable and non-toxic noble gas with high neutron capture cross-section, made it the golden standard for neutron detection, achieving high efficiency and good gamma-ray discrimination. In a proportional counter (PC), $^3$He can work simultaneously as the conversion material, converting neutrons into heavy charged particles, and as the counting gas in which the charged particles deposit their energy. If the detector is large enough, both reaction products deposit their total energy in the proportional counter, and a full energy deposition peak with a Gaussian shape centred at 764 keV is obtained [7].

Despite its excellent properties, an insurmountable drawback is that $^3$He has become extremely scarce [8]. When the scientific community became aware of this shortage, prices sky-rocketed, and heavy acquisition restrictions were implemented. Since $^3$He detectors unfolded as the dominant technology for neutron detection, replacing them on a global scale has not come without tremendous effort. The dimension of this endeavour is such that although more than a decade of intensive research and development of $^3$He free neutron detectors has gone by, about 80% of neutron detectors currently used in neutron scattering applications worldwide are still $^3$He based [9].

## 1.2 $^3$He alternatives and their limitations

One $^3$He free alternative for neutron detection using gaseous detectors relies on the use of solid $^{10}$B, which has a higher thermal neutron capture cross-section than $^6$Li, and consequently the potential to achieve higher detection efficiency. These detectors use a solid layer of a boron-containing material, most commonly natural boron or boron carbide (B$_4$C), which can be $^{10}$B enriched. This material is deposited either on the detector inner walls or on metallic substrates placed inside the detector. The $^{10}$B-neutron capture reaction is composed of two branches:

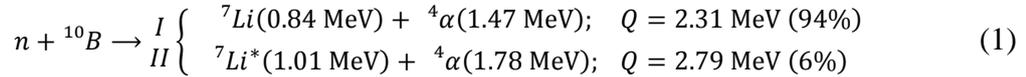

$$n + {}^{10}B \rightarrow \begin{matrix} I \\ II \end{matrix} \begin{cases} {}^7Li(0.84 \text{ MeV}) + {}^4\alpha(1.47 \text{ MeV}); & Q = 2.31 \text{ MeV } (94\%) \\ {}^7Li^*(1.01 \text{ MeV}) + {}^4\alpha(1.78 \text{ MeV}); & Q = 2.79 \text{ MeV } (6\%) \end{cases} \quad (1)$$

Due to momentum conservation, the reaction products are emitted in opposite directions, along the same line (back to back). This geometrical constrain, prevents the detection of both reaction products since only one of them is emitted towards the proportional gas (figure 1-left) while the other is either lost into the $^{10}$B containing layer or its substrate.

It is therefore impossible with these detectors to simultaneously detect the two secondary particles of reaction (1) or to collect the total energy released ($Q$ in equation 1). Instead of a Gaussian peak (as it is the case with $^3$He PC) the pulse height spectrum (PHS) of these detectors is composed by an overlap of 2 continuous distributions, as depicted in Figure 1 (right). Even for the secondary particle of reaction (1) which is emitted into the proportional gas, a fraction of its energy is lost while it crosses the $^{10}$B layer and before it reaches the gas where the remaining energy is deposited via gas ionizations. The fraction of the energy lost depends on the depth at which the neutron is captured, and the direction in which the secondary particles are emitted, giving rise to the 2 steps of the PHS in Figure 1. And since the products of reaction (1) have a range of only a few micrometres in solids, increasing the thickness of the $^{10}$B conversion layer leads to an increase in neutron detection efficiency only up to a certain limit. After that, none of the reaction products can reach the gas and is instead absorbed in the $^{10}$B layer, resulting in undetected neutron captures. The effect of the $^{10}$B converter thickness on the detection efficiency depends on the neutron incoming direction, with the two possible cases (backscattering and transmission) illustrated in Figure 2.

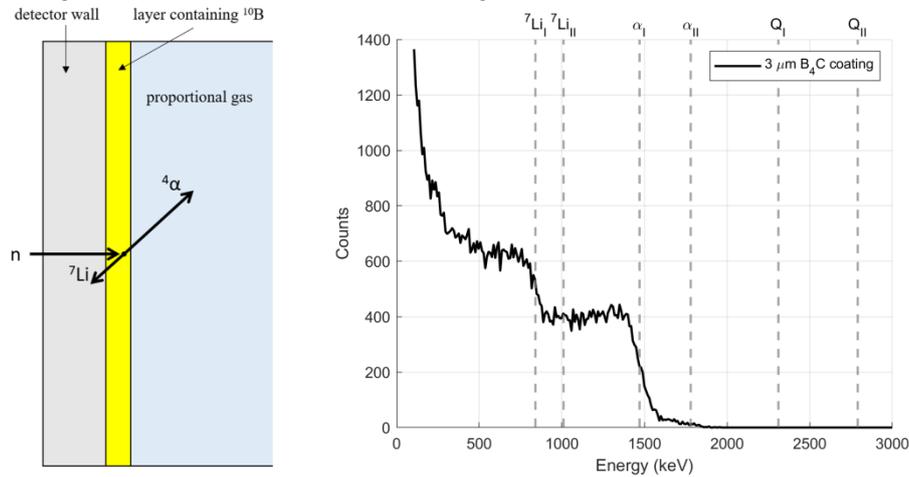

**Figure 1.** Left: Schematics of a neutron capture reaction in a proportional counter having the inner walls coated with a $^{10}$B-containing material. Right: Histogram of the energies depositions in the gas for thermal neutrons interacting with a 3 μm thick $B_4C$ coating deposited on a 0.5 mm thick aluminium foil, obtained by means of a GEANT4 simulation. Detector physical volume was a box with a volume of 10×5×2.5cm$^3$, filled with Ar:$CO_2$ at 1 atm. An energy threshold of 100 keV was applied to the secondaries detected in the physical detector.

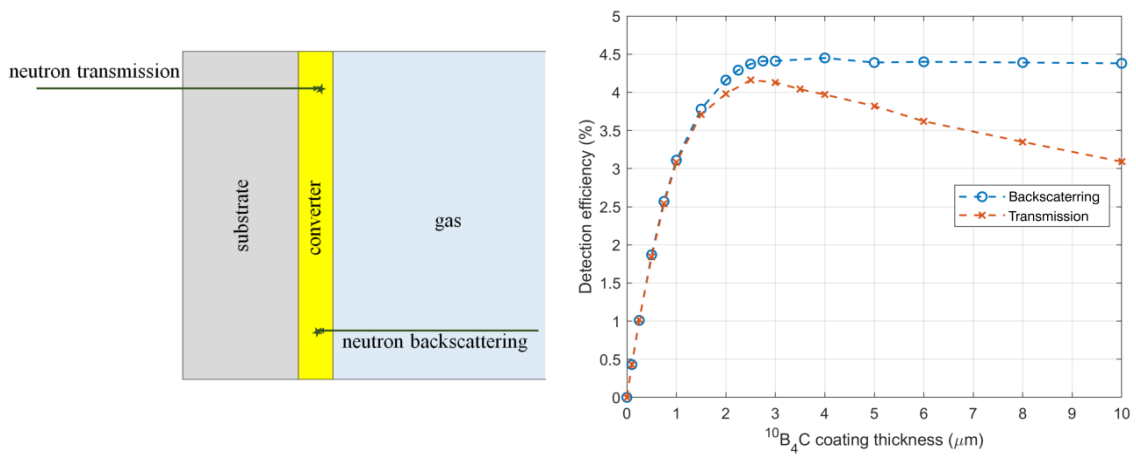

**Figure 2.** Left: layout of neutron interactions in transmission and backscattering modes. Right: Detection efficiency as a function of the $B_4C$ (99% $^{10}$B enriched) thickness, for a neutron beam in transmission and backscattering mode, obtained by GEANT4 simulations. A maximum detection efficiency of ~4.5% is achievable, for a converter thickness of ~3 μm. The simulation geometry was the same as in Figure 1.

– 3 –

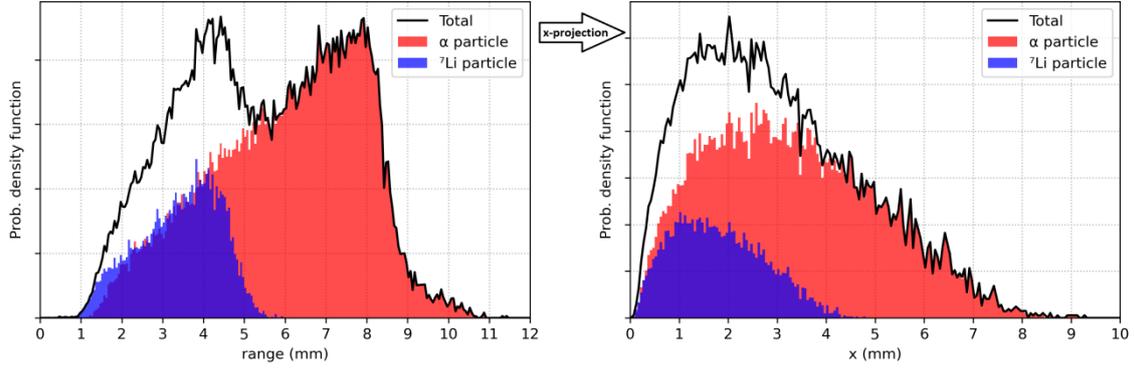

**Figure 3.** Ranges of the $^7$Li and α particle emerging from a 3 μm thick $^{10}$B$_4$C layer into Ar:CO$_2$ (90:10) at atmospheric pressure. The $^7$Li and α particle were produced by a beam of thermal neutrons, perpendicular to the $^{10}$B$_4$C layer, in backscattering mode.

When the $^7$Li or α particle from a neutron capture reaches the gas, it ionizes its atoms, creating a trail of electron-ion pairs. These are formed along the track of the particle in the gas, which can extend to several millimetres at atmospheric pressure. Figure 3 shows the distribution of the ranges of the $^{10}$B neutron reaction products emitted from a 3 μm $^{10}$B$_4$C layer as they enter Ar:CO$_2$ (90:10), a gas commonly used in boron-coated detectors, at atmospheric pressure. An energy cut of 100 keV was applied, used in boron coated neutron detectors to discriminate against gamma rays and electronic noise [10], which nearly resulted in an absence of events for ranges below 1 mm. The results indicate the range of the $^7$Li and α particles can extend from zero to about 11 mm, depending on the depth at which the nuclear capture occurs and the direction in which the particles are emitted. If we consider the projection of the track in a single dimension (Figure 3 - right) the maximum range is approximately 8 mm.

In a gaseous detector the primary electrons resulting from ionizations by the secondary particles in the gas are drifted by an electric field and collected by a readout mechanism, which can provide position information of the particle track. Examples of some readouts are multiwire proportional counters (MWPCs) [10], strip planes [11] or micropatterned gaseous detectors [12-14]. In most cases, several individual readout units are triggered by the $^7$Li and α particles tracks in the gas, which degrades the position resolution. Moreover, the energy lost by the particles along the tracks is not constant, nor is it necessarily greater at the beginning of the track, that is, nearer to the neutron interaction position. In fact, for α particles that escape the conversion layer with a large fraction of their initial energy, the number of ionizations per unit length increases at the end of the track (Figure 4 - left), which adds complexity to position reconstruction. The long particle tracks in the gas intrinsically limit the spatial resolution of position sensitive neutron detectors, being reported as the largest contributor to the position resolution [15].

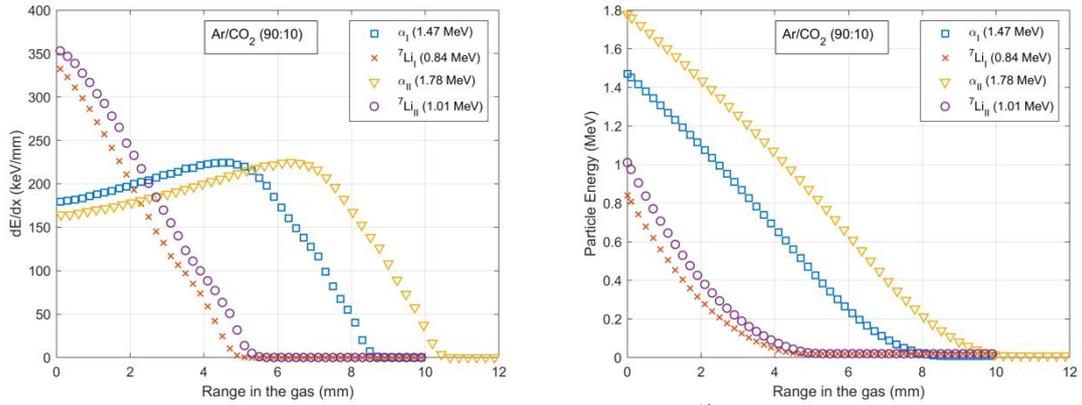

**Figure 4.** Stopping power (left) and kinetic energy (right) of the $^{10}$B decay products as a function of the distance travelled in the gas, in Ar:CO$_2$ (90:10) at atmospheric pressure, obtained with SRIM [16].

## 2. Operation of detectors equipped with thin converter foils and dual readout in coincidence

We propose to improve the position resolution of neutron detectors equipped with $^{10}$B layers, by simultaneous readout of the two products emitted in the neutron capture reaction (1). Since these are emitted back to back, the knowledge of their tracks and the energy deposited by them in the gas will lead to a superior position resolution than the one achieved with detectors than can only sample the position and energy of one of the secondary particles emitted in each neutron capture reaction.

Our method requires the use of a thin layer, containing $^{10}$B atoms, deposited on an also thin substrate, in such a way that both fission fragments from reaction (1) escape the converter foil, ionizing the gas in opposite sides (Figure 5). Two independent readout systems, detectors A and B in Figure 5, are facing each side of the converter foil and detect the two particles emitted in the neutron capture reaction, allowing the reconstruction of the common origin of both tracks. This method makes a better use of the information available in the neutron capture reaction, by collecting the energy and track position of both secondary particles instead of only one of them.

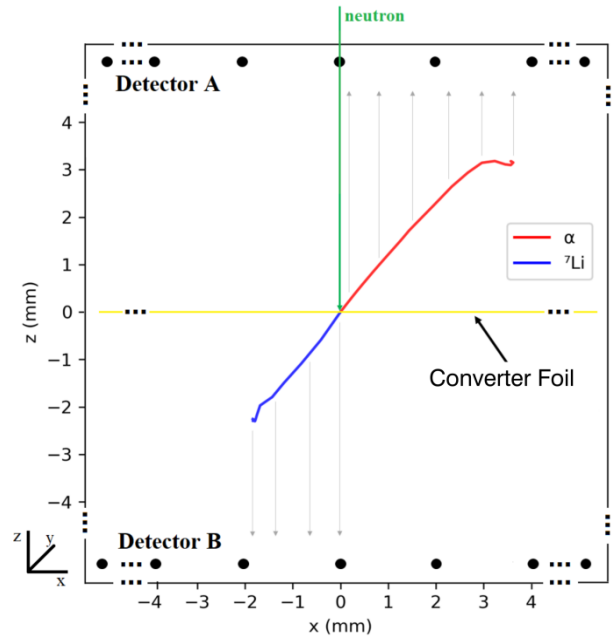

**Figure 5**. GEANT4 simulation of a neutron interacting with a conversion foil made of two B$_4$C layers (0.5 μm each) deposited on both sides of Mylar foil (0.9 μm), in which both secondary particles leave long tracks in the gas.

In the event depicted in Figure 5, the assessment of the neutron capture position on the $^{10}$B layer taking into account only one of the secondary particles is associated with a large uncertainty. Not only the range of the particle in the gas varies according to its energy as it escapes the conversion layer but also the particle is emitted from the converter foil in a random direction. The separate operation of either detector A or detector B is analogous to that of conventional boron-lined detectors, which only detect one fission fragment per neutron capture. Obtaining the neutron capture position for the event depicted in Figure 5 would lay between -2 mm and 0 mm if the information of the $^7$Li ion detected by detector B was used and between 0 mm and 4 mm, for the α particle detected by detector A. However, by combining the information of both detectors it is possible to obtain a value closer to 0 mm, the actual position of the neutron when it interacted with the $^{10}$B converter foil.

Throughout this work we refer to the proposed method as the dual track in coincidence method (2_track), as opposed to the conventional boron coated detectors in which the position information is extracted from just one of the tracks created by the secondary particles (1_track).

## 3. GEANT4 Simulations

In this work we evaluate the feasibility of the proposed method for application to position reconstruction in neutron detectors, by means of GEANT4 [17] simulations. In the geometry implemented the converter foil is made of a 0.9 μm Mylar film, with one layer of B$_4$C deposited on each side of the foil. Although in some cases the $^{10}$B containing material can be self-suporting, not requiring the use of a substrate where it is deposited [18] these solutions are limited to small area coverage, which can be an issue for imaging detectors. Hence, in our simulations we will consider the use of a substrate in which the $^{10}$B containing material is deposited, allowing for larger area coverage. We found that Mylar films are a suitable option, due to their easiness in the manipulation process and excellent mechanical properties. In our simulations, a 0.9 μm Mylar film was selected as support for the $^{10}$B containing material. These films are commercially available in thicknesses down to 0.5 μm, and therefore the choice of 0.9 μm can be seen as a conservative one. The choice for the material containing the $^{10}$B atoms was B$_4$C, $^{10}$B enriched to 99%, which is commonly used in neutron detectors and can be deposited on Mylar films by several methods. The total thickness of B$_4$C was divided in equal parts by the two sides of the Mylar film, resulting in a symmetric arrangement. This has the advantage of not leaving the Mylar film exposed to the proportional gas, which could lead to static charge build up during detector operation under high gains and/or high rate. To optimize the detector geometry, the relationship between total thickness of the $^{10}$B$_4$C layer and the detection efficiency was evaluated by means of GEANT4 simulations. The selected physics list was the QGSP_BERT_HP, recommended for the transport of neutrons down to thermal energies [19], with range cut of 50 nm for the solid materials and 10 μm in the gas. The converter foil was placed in the centre of the geometry, with two sensitive detectors made of Ar:CO$_2$ (90:10) at atmospheric pressure placed on its sides. The foil had an area of $10 \times 10$ cm$^2$ and each sensitive detector extended over a depth of 5 cm. For each simulation, a total of $10^6$ thermal neutrons were produced as primary particles. These were generated at x= 0 in our geometry, as depicted in the event of Figure 5, and constitute an infinitesimal small neutron beam, emitted towards the converter foil, in a direction perpendicular to it and aiming at its centre. The simulation outputs the energy deposited along each step in the sensitive detectors and the corresponding position. An energy cut of 100 keV was applied to the outputs of the simulations, removing events in which the energy deposited in the gas by any of the secondary particles was inferior to this threshold.

For each detector, A and B in our simulations, the position of the neutron beam in the converter foil along the x-coordinate was reconstructed by averaging the x-coordinate of each discrete step along the secondary particle trajectory in the gas volume, weighted by the energy deposited in the corresponding step:

$$\hat{x}_{1\_track} = \frac{\sum_i x_i dE_i}{\sum_i dE_i} \qquad (2)$$

For the application of the dual track in coincidence method, we select events in which there was a deposition of energy on both detectors A and B. The x-coordinates in these detectors, obtained from equation (2), are then averaged by weighting over the total energy deposited in the opposite detector:

$$\hat{x}_{2\_track} = \frac{\hat{x}_{A_{1\_track}} \times E_B + \hat{x}_{B_{1\_track}} \times E_A}{E_A + E_B} \qquad (3)$$

This weighting yields superior results than the simple average between $\hat{x}_A$ and $\hat{x}_B$, since it assures that the secondary particle that carries more energy (and consequently produces a longer electron-ion track in the gas) has less weight than the particle with a shorter track, returning a neutron capture site closer to the real one.

The histogram of the events reconstructed using equations (2) and (3) is fitted to Gaussian curve, from which the neutron position on the converter foil and the position resolution are taken from the mean and FWHM of the fit, respectively.

## 4. Results

### 4.1 Pulse height spectrum and detection efficiency

The pulse height spectrum of the events resulting from the 2_track method (Figure 6, left) is obtained by summing the energies in detectors A and B for events that produced an energy deposition above the 100 keV threshold on both detectors. It has therefore a very different shape than that of detectors A and B, resembling the one of $^3$He filled proportional counters, in which the energy of neutron capture reaction products can be fully collected. However, in this case, the main peak is not centred at the total reaction energy ($Q_I$ = 2310 keV), due to the inevitable energy losses by the secondary particles while travelling in the $B_4C$ and Mylar of the converter foil. As a result, the peak is shifted towards lower energies, with its centroid around 1450 keV (for a total $B_4C$ thickness of 1 μm) and does not have a Gaussian shape, but rather a tail down to lower energies, due to the contributions of neutron captures in which the secondary particles lose a significant fraction of their energy before reaching the gas. A smaller peak is also visible around 1950 keV, from the $^{10}$B neutron captures in the 6% probability branch in reaction 1 ($Q_{II}$ = 2790 keV).

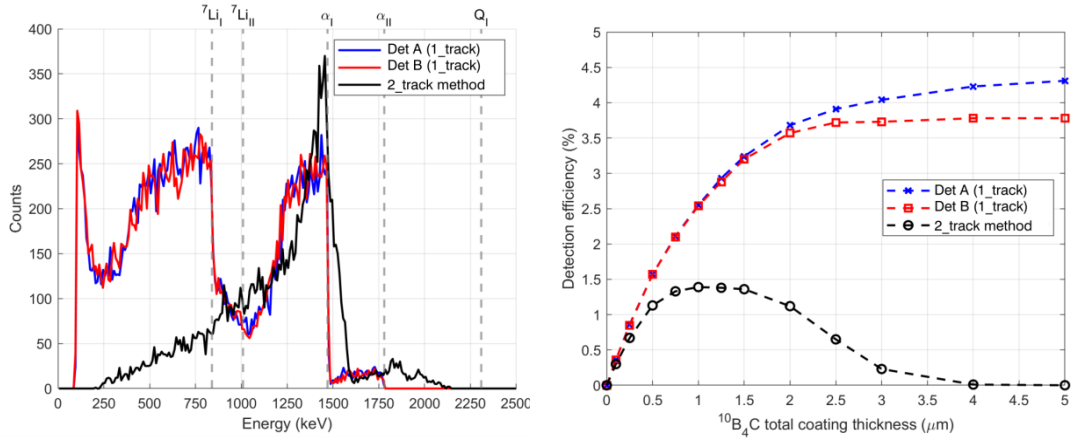

**Figure 6**. Results of GEANT4 simulations of 0.9 μm Mylar film covered on both sides with $B_4C$ (99% $^{10}$B enriched) layers having equal thickness. Left: Pulse heigh spectrum, obtained for detector A, detector B and with the 2_track method for a $B_4C$ total thickness of 1 μm. Right: Detection efficiency for detectors A and B, and the one resulting from application of the 2_track method, as a function of the $B_4C$ total thickness. Detectors A and B positions relative to the incoming neutron beam are the same as depicted in Figure 5.

The 2_track method only accounts for neutron captures that simultaneously deposit energy in both gas regions. Therefore, its efficiency is not the sum of the efficiencies of detectors A and B, since in some neutron captures only one of the secondary particles reaches the gas. A $B_4C$ total thickness of 1 μm was found to maximize the detection efficiency, for which the value was 1.4% (Figure 6, left), and was selected for application of the 2_track method. Despite this modest detection efficiency, the type of information that can be obtained with the 2_track method allows for a superior position reconstruction of the incoming neutrons, even if a relatively low number of events are used for the reconstruction, as will be discussed in the next section. The 2_track method can also benefit from traditional approaches known to result in large detection efficiency improvements, such as tilting the converter and cascading several individual elements [19].

**4.2 Spatial resolution**

The distribution of the positions reconstructed using the 2_track method (equation 3), for a converter foil made of a 0.9 μm Mylar film, covered on each side with a 0.5 μm layer of $B_4C$, is plotted in figure 7. It is compared with the one obtained using equation 2 (1_track method), for a 1 μm thick $B_4C$ layer (Figure 7, left) and for a 3 μm thick $B_4C$ layer (Figure 7, right). The former corresponds to the same total $B_4C$ thickness as the one selected for the 2_track method, while the latter corresponds to the thickness for which the maximum detection efficiency is achieved, for detectors operating in backscattering mode.

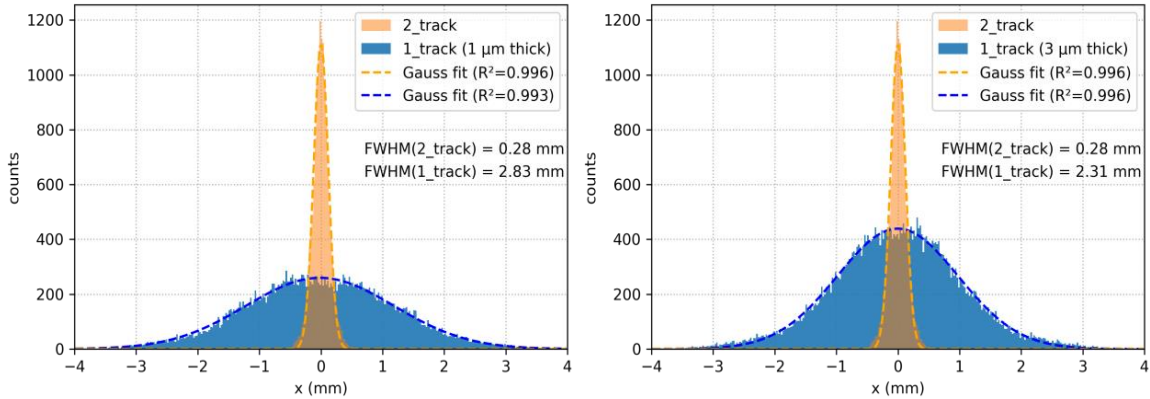

**Figure 7.** Distribution of the reconstructed positions with the 2_track method, for a 1 μm $B_4C$ total thickness on both sides of a 0.9 μm Mylar foil (orange). Comparison with the one of the 1_track method (blue) for a 1 μm (left) and a 3 μm (right) thick $B_4C$ layer. A total of $10^6$ primary particles were generated for each simulation which resulted in ~$10^4$ detected events for the 2_track method and ~$4\times10^4$ for the 3 μm thick 1_track method.

From Figure 7 we can observe the notable spatial resolution improvement with the 2_track method: while a conventional $B_4C$ coated detector (1_track) with a 3 μm thick $B_4C$ layer is associated with a FWHM = 2.31 mm, the 2_track method can provide a FWHM = 0.28 mm, which represents an improvement by a factor of 8.25. Figure 7 clearly demonstrates the potential in the 2_track method, and the superior use it makes of the information available in the neutron capture reaction (line of action and energies of the secondary particles).

An additional attractive feature of the 2_track method is that the reconstructed position converges significantly faster to the correct value, as shown in Figure 8: not only the ultimate position resolution with this method is better than the one of conventional $B_4C$ coated detectors, its ability to accurately reconstruct the position is also superior even for lower exposures. As an example, the reconstructed positions (Figure 8, left) with the 2_track method are within ±0.05 mm already for $10^2$ detected events. To obtain the same precision with the 1_track detector (3 μm thick), around $2\times10^3$ detected events are required. Even if we take into account the differences in detection efficiency (1.4% and 4.5 % for the 2_track and 1_track with a 3 μm layer, respectively), the 2_track method is associated with a reduction of the exposure time by a factor of 6. Hence, it presents a far superior position resolution and better precision, even for shorter exposure times.

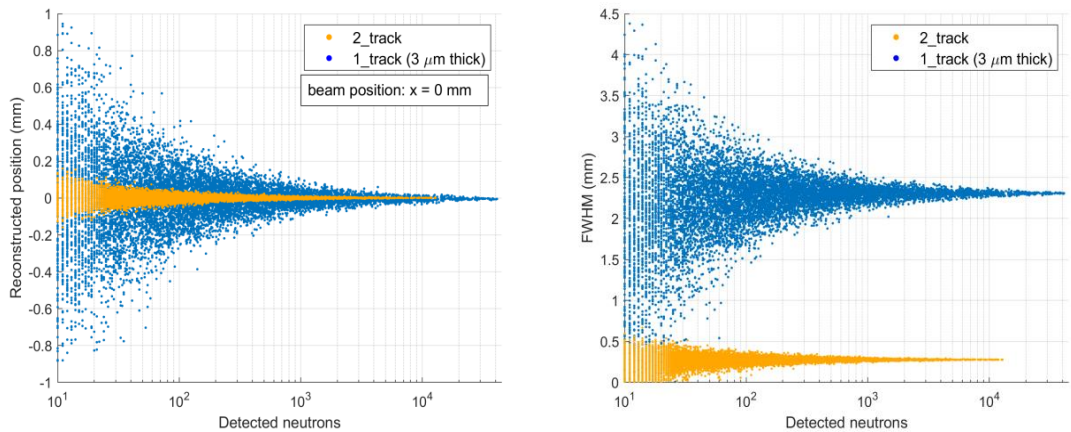

**Figure 8.** Reconstructed neutron capture x-coordinate (left) and FWHM (right) as function of the number of events used in the Gaussian fits of figures 7, for the 1_track detector with a 3 μm thick $B_4C$ layer (blue) and for the 2_track method (orange).

**5. Conclusions and future work**

A novel concept for slow neutron detection is presented, that aims to improve the intrinsic spatial resolution of gaseous neutron detectors with solid boron converters, by simultaneously detecting both reaction products emitted in the $^{10}B$ neutron capture reaction. Because the $^7Li$ and α particles are emitted back to back, by tracking both particles it is possible to pinpoint the neutron interaction position. This requires that both particles escape the solid boron layer with a significant fraction of their initial energy, which is possible if the combined thickness of the boron layer and required supporting material is lower than the ranges of the secondary particles in these materials. In this work we have introduced this novel concept and, by means of simulations with the GEANT4 software package, optimized the converter thickness and presented the first results regarding the potential improvements in position resolution.

A maximum detection efficiency of 1.4% was obtained, from simulations of a thermal neutron beam perpendicular to a converter foil made by a 0.9 μm thick Mylar film with a 0.5 μm $B_4C$ layer ($^{10}B$ enriched to 99%) on each side. Results of the simulations show an improvement by a factor of 8 in spatial resolution and allow inferring a considerable reduction in the exposure time when compared to conventional detectors.

Although the absolute detection efficiency resulting from application of the dual track in coincidence method is inferior to that of a conventional detector, our results show that it can generate a more precise neutron position information, even for a substantially lower number of detected events. The lower absolute detection efficiency may also allow the exposure to higher neutron fluxes before reaching its counting rate limit. In addition, the detection concept we present is compatible with the conventional strategies used to improve detection efficiency: stacking consecutive detection layers, tilting the detector to form a grazing angle with the incoming neutrons, or a combination of both [10].

Our results take into account only the contribution of the range of the secondary particles to the spatial resolution. Although it is considered to be the main contributor to position resolution [15], it is important to include other contributions in future work, such as those caused by charge transfer and collection processes in the gaseous detector, in order to reach a more accurate spatial resolution. The geometry of the gaseous detector, namely the depth of the drift region, also impacts significantly the position resolution [21]. The effect of these contributions can be measured using the spatial distribution of the primary charges obtained with GEANT4 as an input for simulation software such as Garfield++ [22], which provides a detailed simulation of two and three-dimensional drift chambers, where geometry and polarization of the readout electrodes can be defined, to generate a corresponding detector response.

Several options are being considered to produce the converter foil. 1 μm thick boron foils are commercially available in the form of discs with 1 cm diameter [18]. Such thickness is suitable for immediate application and experimental verification of this novel detection concept. Depositing $B_4C$ over larger areas (100 cm$^2$) can be performed by magnetron sputtering techniques, with the use of a Mylar (or other material) film as support. Efforts in that direction are on-going at the Paul Scherrer Institute, with promising results already achieved, although the deformation of the thermo-sensitive Mylar foil after the sputtering process has yet to be solved. Other concerns regarding converter foil production, namely the non-stoichiometric deposition and the mechanical properties of the film, motivate us to pursue other production methods such as coatings made by dripping a solution containing $B_4C$ nano-particles [23] or room temperature pulsed laser deposition [24,25], both under consideration for the production of large area ultra-thin neutron converter foils for application to this novel technique.

## Acknowledgments

Support is acknowledged under research projects PTDC/NAN-MAT/30178/2017, CERN/FIS-INS/0026/2019, IF/00039/2015 and UIDP/04559/2020, funded by national funds through FCT/MCTES and co-financed by the European Regional Development Fund (ERDF) through the Portuguese Operational Program for Competitiveness and Internationalization, COMPETE 2020. N. Duarte acknowledges the support of FCT, under contract PD/BD/128268/2016.